\documentclass{article}
\usepackage{graphicx} 
\usepackage{amsfonts}
\usepackage{amsmath}
\usepackage{amsthm}
\usepackage{booktabs}
\usepackage{url}
\usepackage{etoolbox}
\usepackage{tabularx}
\usepackage{adjustbox}
\usepackage{hyperref}
\usepackage{authblk}
\usepackage[yyyymmdd]{datetime}

\newcommand{\GV}{Greedy Vanilla}
\newcommand{\GP}{Greedy Potential}

\newcommand{\tabref}[1]{Table~\ref{tab:#1}}

\title{A Rank 23 Algorithm for Multiplying $3 \times 3$ Matrices with an Arithmetic Complexity of 59}

\author[1,2]{Erik Mårtensson}
\author[1]{Paul Stankovski Wagner}
\author[3]{Joshua Stapleton}

\date{}
\affil[1]{Lund University, Lund, Sweden \texttt{{\small \{erik.martensson,paul.stankovski\_wagner\}@eit.lth.se}}}
\affil[2]{Advenica AB, Malm\"o, Sweden}
\affil[3]{Department of Mathematics, Imperial College London, UK \texttt{{\small jbs123@ic.ac.uk}}}

\begin{document}

\maketitle

\begin{abstract}
In 1969 Strassen showed surprisingly that it is possible to multiply two $2 \times 2$ matrices using seven multiplications and 18 additions, instead of the naive eight multiplications and four additions. The number of additions was later reduced to 15. Karstadt and Schwartz further reduced the number of additions to 12 using a change-of-basis method. Both the number of multiplications and the number of additions have been shown to be optimal for the $2 \times 2$ case.

For multiplying $3 \times 3$ matrices, the lowest number of multiplications found so far is 23. Using 23 multiplications, Schwart et al. showed how to reduce the number of additions to 61 using a change-of-basis method. Mårtensson and Stankovski Wagner showed how to achieve 62 additions, without changing basis. Using the optimization method by Mårtensson and Stankovski Wagner, Stapleton found an algorithm requiring only 60 additions. In this work we continue to combine the methods of Mårtensson, Stankovski Wagner and Stapleton, finding an algorithm requiring only 59 additions, still without a basis change. Technical details on the method and tools used for finding this scheme, and a discussion on the impact of this discovery, will come in an upcoming publication.
\end{abstract}

\section{Introduction}
Multiplying $2 \times 2$ matrices directly from the definition requires eight multiplications and four additions. In 1969 Strassen~\cite{Strassen} showed how to reduce the number of multiplications to seven, at the cost of increasing the number of additions to 18\footnote{When referring to additions, we also implicitly include both additions and subtractions.}. The number of multiplications is also called the rank of the algorithm. The rank seven was shown to be optimal in~\cite{Winograd}.

For rank 7 algorithms for multiplying $2 \times 2$ matrices, the number of additions was reduced to 15 in~\cite{15Optimal}. The number of additions required is also referred to as the arithmetic complexity of the algorithm. Using a change of basis, Karstadt and Schwartz improved this number to 12 and showed it to be optimal in~\cite{KarstadtSchwartz}. In terms of computational complexity, performing this change of basis introduces lower-order terms that are small compared to the cost of performing an addition.

Now consider multiplying an $n \times m$ matrix by an $m \times k$ matrix. For most of the matrix dimensions larger than that of $2 \times 2$, both the optimal rank and the optimal arithmetic complexity of matrix multiplication are not known. For a collection of the schemes found for a large number of dimensions, see~\cite{sedoglavicBible}.

For multiplying $3 \times 3$ matrices, Laderman found a rank 23 algorithm by hand in 1976~\cite{Laderman}. After almost half a century of applying increasingly sophisticated search methods, this rank has not been improved. In this work, we introduce a new record for the number of additions for a rank 23 algorithm for multiplying $3 \times 3$ matrices.

In terms of the number of additions required for rank 23 algorithms for multiplying $3 \times 3$ matrices, Smirnov found an algorithm that naively requires 84 additions~\cite{Smirnov}. Smirnov pointed out that the number of additions for his algorithm can be lowered, but he did not perform such an optimization. Karstadt and Schwartz reduced the number of additions to 75 in~\cite{KarstadtSchwartz}. Beniamini et al. reduced this number further to 61 in~\cite{beniamini61Additions}. Without performing a change of basis, Mårtensson and Stankovski Wagner showed how to perform Laderman's original algorithm using only 62 additions, almost matching the result of Beniamini et al.~\cite{MartenssonStankovski}.

Recently, Stapleton developed a method for generating low-rank matrix multiplication schemes using a neural network~\cite{Stapleton}. By generating many schemes of rank 23 and reducing the number of additions using the implementations of Mårtensson and Stankovski Wagner, he found a 60-addition algorithm.

We further combined Stapleton's method to generate fast matrix multiplication schemes with Mårtensson and Stankovski Wagner's method for reducing the number of additions to find a scheme requiring only 59 additions. The naive (unreduced) form of the scheme uses 110 additions and is specified in~\tabref{stapleton_59_before}. Our optimized scheme is specified in~\tabref{stapleton_59_after}, requiring only 59 additions. We use the compact indexing notation
\[
\resizebox{.9\hsize}{!}{$\begin{pmatrix}
        C_{0} & C_{1}  & C_{2}\\
        C_{3} & C_{4}  & C_{5}\\
        C_{6} & C_{7}  & C_{8}
    \end{pmatrix} = \mathbf{C} = \mathbf{A}\mathbf{B} = \begin{pmatrix}
        A_{0} & A_{1}  & A_{2}\\
        A_{3} & A_{4}  & A_{5}\\
        A_{6} & A_{7}  & A_{8}
    \end{pmatrix}
    \begin{pmatrix}
        B_{0} & B_{1} & B_{2}\\
        B_{3} & B_{4} & B_{5}\\
        B_{6} & B_{7} & B_{8}
    \end{pmatrix}$.}
\]

Finally, for convenience, in the Appendix we have specified the 59-algorithm on a form readable by the the software implementations of Mårtensson~\cite{github-erik} and Stankovski Wagner~\cite{github-paul} in ~\tabref{fileFormat}.

More technical details on how the scheme was found, tools and methods used, as well as implications of the finding, will be included in an upcoming publication.

\begin{table*}[ht!]
    \centering
\begin{adjustbox}{max width=.99\textwidth}
\begin{tabularx}{\textwidth}{X}
\small
\begin{eqnarray*}
M_{0} &=& \phantom{(}A_{1}B_{4}\\
M_{1} &=& \phantom{(}A_{0}B_{1}\\
M_{2} &=& (A_{3}-A_{4}-A_{5}+A_{6}-A_{7}-A_{8})B_{8}\\
M_{3} &=& (A_{2}-A_{8})B_{8}\\
M_{4} &=& (A_{3}-A_{5}+A_{6}-A_{8})B_{6}\\
M_{5} &=& \hphantom{(}A_{2}B_{7}\\
M_{6} &=& \hphantom{(}A_{0}B_{0}\\
M_{7} &=& (A_{1}-A_{7})(B_{2}+B_{5})\\
M_{8} &=& \phantom{(}A_{5}(B_{4}+B_{5}-B_{6}-B_{7}-B_{8})\\
M_{9} &=& \phantom{(}A_{4}B_{3}\\
M_{10} &=& (A_{0}-A_{1}-A_{6}+A_{7})B_{2}\\
M_{11} &=& (A_{3}+A_{6}-A_{8})(B_{1}+B_{2}+B_{4}+B_{5}-B_{6})\\
M_{12} &=& (A_{1}+A_{6}-A_{7})(B_{2}+B_{4}+B_{5})\\
M_{13} &=& \phantom{(}A_{1}B_{3}\\
M_{14} &=& (A_{1}+A_{3}+A_{6}-A_{7})(B_{2}-B_{3}+B_{5})\\
M_{15} &=& (A_{3}-A_{4}+A_{6}-A_{7})(B_{3}-B_{5}+B_{8})\\
M_{16} &=& (A_{3}-A_{4}-A_{5}+A_{6}-A_{7})(B_{4}+B_{5}-B_{8})\\
M_{17} &=& \phantom{(}A_{3}(B_{0}+B_{1}+B_{2}+B_{4}+B_{5})\\
M_{18} &=& (A_{3}-A_{4}-A_{5})(B_{4}+B_{5})\\
M_{19} &=& \phantom{(}A_{8}(B_{1}+B_{2}+B_{4}+B_{5}+B_{7})\\
M_{20} &=& \phantom{(}A_{2}B_{6}\\
M_{21} &=& (A_{6}-A_{8})(B_{1}+B_{2}+B_{4}+B_{5})\\
M_{22} &=& (A_{3}+A_{6})(B_{0}-B_{2}+B_{3}-B_{5}+B_{6})\\
~\\
C_{0} &=& \phantom{-}M_{6}+M_{13}+M_{20}\\
C_{1} &=& \phantom{-}M_{0}+M_{1}+M_{5}\\
C_{2} &=& -M_{0}-M_{2}+M_{3}+M_{10}+M_{12}-M_{16}+M_{18}\\
C_{3} &=& -M_{4}+M_{9}-M_{11}+M_{17}+M_{21}\\
C_{4} &=& -M_{0}+M_{4}-M_{8}-M_{9}+M_{11}+M_{12}-M_{13}-M_{14}-M_{15}-M_{16}-M_{21}\\
C_{5} &=& \phantom{-}M_{0}+M_{9}-M_{12}+M_{13}+M_{14}+M_{15}+M_{16}-M_{18}\\
C_{6} &=& -M_{7}+M_{11}+M_{13}+M_{14}-M_{17}-M_{21}+M_{22}\\
C_{7} &=& \phantom{-}M_{0}+M_{7}-M_{12}+M_{19}+M_{21}\\
C_{8} &=& -M_{0}-M_{2}-M_{7}+M_{12}-M_{16}+M_{18}
\end{eqnarray*}
\end{tabularx}
\end{adjustbox}
    \caption{Fast matrix multiplication algorithm with 110 additions (before reduction). Here we use compact indexing notation. The algorithm is specified in implementation order. Note that negation (negative leading term) is entirely avoided in implementations by rearranging terms.}
    \label{tab:stapleton_59_before}
\end{table*}

\begin{table}[ht!]
    \centering
\begin{adjustbox}{max width=.99\textwidth}
\begin{tabularx}{\textwidth}{>{\hsize.5\hsize\raggedright\arraybackslash}X>{\hsize.5\hsize}X}
\small
\begin{eqnarray*}
t_{0} &=& A_{3}+A_{6}\\
t_{1} &=& A_{1}-A_{7}\\
t_{2} &=& A_{4}+A_{5}\\
t_{3} &=& A_{7}-t_{0}\\
t_{4} &=& A_{6}+t_{1}\\
t_{5} &=& A_{8}-t_{0}\\
t_{6} &=& t_{2}+t_{3}\\
~\\
u_{0} &=& B_{2}+B_{5}\\
u_{1} &=& B_{4}+u_{0}\\
u_{2} &=& B_{1}+u_{1}\\
u_{3} &=& B_{4}+B_{5}\\
u_{4} &=& B_{3}-u_{0}\\
u_{5} &=& B_{8}-u_{3}\\
~\\
M_{0} &=& \phantom{(}A_{1}B_{4}\\
M_{1} &=& \phantom{(}A_{0}B_{1}\\
M_{2} &=& (A_{8}+t_{6})B_{8}\\
M_{3} &=& (A_{2}-A_{8})B_{8}\\
M_{4} &=& (A_{5}+t_{5})B_{6}\\
M_{5} &=& \phantom{(}A_{2}B_{7}\\
M_{6} &=& \phantom{(}A_{0}B_{0}\\
M_{7} &=& \phantom{(}t_{1}u_{0}\\
M_{8} &=& \phantom{(}A_{5}(B_{6}+B_{7}+u_{5})\\
M_{9} &=& \phantom{(}A_{4}B_{3}\\
M_{10} &=& (A_{0}-t_{4})B_{2}\\
M_{11} &=& \phantom{(}t_{5}(B_{6}-u_{2})\\
M_{12} &=& \phantom{(}t_{4}u_{1}\\
M_{13} &=& A_{1}B_{3}
\end{eqnarray*}
&
\small
\begin{eqnarray*}
M_{14} &=& (t_{0}+t_{1})u_{4}\\
M_{15} &=& (A_{4}+t_{3})(B_{3}-B_{5}+B_{8})\\
M_{16} &=& \phantom{(}t_{6}u_{5}\\
M_{17} &=& \phantom{(}A_{3}(B_{0}+u_{2})\\
M_{18} &=& (A_{3}-t_{2})u_{3}\\
M_{19} &=& \phantom{(}A_{8}(B_{7}+u_{2})\\
M_{20} &=& \phantom{(}A_{2}B_{6}\\
M_{21} &=& (A_{6}-A_{8})u_{2}\\
M_{22} &=& \phantom{(}t_{0}(B_{0}+B_{6}+u_{4})\\
~\\
v_{0} &=& \phantom{-}M_{0}-M_{12}\\
v_{1} &=& \phantom{-}M_{16}+v_{0}\\
v_{2} &=& \phantom{-}M_{11}-M_{21}\\
v_{3} &=& \phantom{-}M_{14}-M_{13}\\
v_{4} &=& \phantom{-}M_{18}-v_{1}\\
v_{5} &=& \phantom{-}M_{2}+v_{4}\\
v_{6} &=& \phantom{-}M_{4}+M_{9}\\
v_{7} &=& \phantom{-}M_{15}+v_{3}\\
v_{8} &=& \phantom{-}M_{17}-v_{2}\\
~\\
C_{0} &=& \phantom{-}M_{6}+M_{13}+M_{20}\\
C_{1} &=& \phantom{-}M_{0}+M_{1}+M_{5}\\
C_{2} &=& \phantom{-}M_{3}+M_{10}+v_{5}\\
C_{3} &=& \phantom{-}v_{6}+v_{8}\\
C_{4} &=& \phantom{-}M_{8}-v_{1}+v_{2}-v_{6}+v_{7}\\
C_{5} &=& \phantom{-}M_{9}-v_{4}-v_{7}\\
C_{6} &=& -M_{7}+M_{22}-v_{3}-v_{8}\\
C_{7} &=& \phantom{-}M_{7}+M_{19}+M_{21}+v_{0}\\
C_{8} &=& -M_{7}+v_{5}
\end{eqnarray*}
\end{tabularx}
\end{adjustbox}
    \caption{Fast matrix multiplication algorithm reduced to 59 additions with \GV{} from~\cite{MartenssonStankovski}. \GP{} yields no improvement. Here we use compact indexing notation. The algorithm is specified in implementation order. Note that negation (negative leading term) is entirely avoided in implementations by rearranging terms.}
    \label{tab:stapleton_59_after}
\end{table}

\clearpage

\bibliographystyle{alpha}
\bibliography{References}

\newpage

\begin{appendix}


\section{The 59-algorithm in File Format}
    For convenience and ease of use, we include our 59-algorithm in file format too.

\begin{table}[ht!]
\begin{adjustbox}{max width=.99\textwidth}
    \begin{tabular}{ccccc ccccc ccccc ccccc ccc}
    \setlength{\tabcolsep}{6pt}
         0 & 1 & 0 & 0 & 0 & 0 & 1 & 0 & 0 & 0 & 1 & 0 & 0 & 0 & 0 & 0 & 0 & 0 & 0 & 0 & 0 & 0 & 0\\
1 & 0 & 0 & 0 & 0 & 0 & 0 & -1 & 0 & 0 & -1 & 0 & 1 & -1 & 1 & 0 & 0 & 0 & 0 & 0 & 0 & 0 & 0\\
0 & 0 & 0 & 1 & 0 & 1 & 0 & 0 & 0 & 0 & 0 & 0 & 0 & 0 & 0 & 0 & 0 & 0 & 0 & 0 & -1 & 0 & 0\\
0 & 0 & 1 & 0 & 1 & 0 & 0 & 0 & 0 & 0 & 0 & -1 & 0 & 0 & 1 & 1 & 1 & -1 & 1 & 0 & 0 & 0 & -1\\
0 & 0 & -1 & 0 & 0 & 0 & 0 & 0 & 0 & 1 & 0 & 0 & 0 & 0 & 0 & -1 & -1 & 0 & -1 & 0 & 0 & 0 & 0\\
0 & 0 & -1 & 0 & -1 & 0 & 0 & 0 & 1 & 0 & 0 & 0 & 0 & 0 & 0 & 0 & -1 & 0 & -1 & 0 & 0 & 0 & 0\\
0 & 0 & 1 & 0 & 1 & 0 & 0 & 0 & 0 & 0 & -1 & -1 & 1 & 0 & 1 & 1 & 1 & 0 & 0 & 0 & 0 & -1 & -1\\
0 & 0 & -1 & 0 & 0 & 0 & 0 & 1 & 0 & 0 & 1 & 0 & -1 & 0 & -1 & -1 & -1 & 0 & 0 & 0 & 0 & 0 & 0\\
0 & 0 & -1 & -1 & -1 & 0 & 0 & 0 & 0 & 0 & 0 & 1 & 0 & 0 & 0 & 0 & 0 & 0 & 0 & -1 & 0 & 1 & 0\\
\#\\
0 & 0 & 0 & 0 & 0 & 0 & 1 & 0 & 0 & 0 & 0 & 0 & 0 & 0 & 0 & 0 & 0 & 1 & 0 & 0 & 0 & 0 & 1\\
0 & 1 & 0 & 0 & 0 & 0 & 0 & 0 & 0 & 0 & 0 & -1 & 0 & 0 & 0 & 0 & 0 & 1 & 0 & -1 & 0 & -1 & 0\\
0 & 0 & 0 & 0 & 0 & 0 & 0 & -1 & 0 & 0 & -1 & -1 & 1 & 0 & 1 & 0 & 0 & 1 & 0 & -1 & 0 & -1 & -1\\
0 & 0 & 0 & 0 & 0 & 0 & 0 & 0 & 0 & 1 & 0 & 0 & 0 & 1 & -1 & -1 & 0 & 0 & 0 & 0 & 0 & 0 & 1\\
1 & 0 & 0 & 0 & 0 & 0 & 0 & 0 & 1 & 0 & 0 & -1 & 1 & 0 & 0 & 0 & 1 & 1 & -1 & -1 & 0 & -1 & 0\\
0 & 0 & 0 & 0 & 0 & 0 & 0 & -1 & 1 & 0 & 0 & -1 & 1 & 0 & 1 & 1 & 1 & 1 & -1 & -1 & 0 & -1 & -1\\
0 & 0 & 0 & 0 & 1 & 0 & 0 & 0 & -1 & 0 & 0 & 1 & 0 & 0 & 0 & 0 & 0 & 0 & 0 & 0 & -1 & 0 & 1\\
0 & 0 & 0 & 0 & 0 & 1 & 0 & 0 & -1 & 0 & 0 & 0 & 0 & 0 & 0 & 0 & 0 & 0 & 0 & -1 & 0 & 0 & 0\\
0 & 0 & 1 & 1 & 0 & 0 & 0 & 0 & -1 & 0 & 0 & 0 & 0 & 0 & 0 & -1 & -1 & 0 & 0 & 0 & 0 & 0 & 0\\
\#\\
0 & 0 & 0 & 0 & 0 & 0 & 1 & 0 & 0 & 0 & 0 & 0 & 0 & -1 & 0 & 0 & 0 & 0 & 0 & 0 & 1 & 0 & 0\\
1 & 1 & 0 & 0 & 0 & 1 & 0 & 0 & 0 & 0 & 0 & 0 & 0 & 0 & 0 & 0 & 0 & 0 & 0 & 0 & 0 & 0 & 0\\
-1 & 0 & -1 & 1 & 0 & 0 & 0 & 0 & 0 & 0 & -1 & 0 & 1 & 0 & 0 & 0 & -1 & 0 & -1 & 0 & 0 & 0 & 0\\
0 & 0 & 0 & 0 & -1 & 0 & 0 & 0 & 0 & 1 & 0 & -1 & 0 & 0 & 0 & 0 & 0 & -1 & 0 & 0 & 0 & 1 & 0\\
-1 & 0 & 0 & 0 & 1 & 0 & 0 & 0 & -1 & -1 & 0 & 1 & 1 & 1 & -1 & 1 & -1 & 0 & 0 & 0 & 0 & -1 & 0\\
1 & 0 & 0 & 0 & 0 & 0 & 0 & 0 & 0 & 1 & 0 & 0 & -1 & -1 & 1 & -1 & 1 & 0 & 1 & 0 & 0 & 0 & 0\\
0 & 0 & 0 & 0 & 0 & 0 & 0 & -1 & 0 & 0 & 0 & 1 & 0 & -1 & 1 & 0 & 0 & 1 & 0 & 0 & 0 & -1 & -1\\
1 & 0 & 0 & 0 & 0 & 0 & 0 & 1 & 0 & 0 & 0 & 0 & -1 & 0 & 0 & 0 & 0 & 0 & 0 & 1 & 0 & 1 & 0\\
-1 & 0 & -1 & 0 & 0 & 0 & 0 & -1 & 0 & 0 & 0 & 0 & 1 & 0 & 0 & 0 & -1 & 0 & -1 & 0 & 0 & 0 & 0
    \end{tabular}
\end{adjustbox}
  \caption{The 59-algorithm written in a format that is readable by the implementations of Mårtensson~\cite{github-erik} and Stankovski Wagner~\cite{github-paul}.}
  \label{tab:fileFormat}
\end{table}

\end{appendix}

\end{document}